\documentclass{Interspeech2024}

\interspeechcameraready

\usepackage{times}
\usepackage{latexsym}
\usepackage[T1]{fontenc}
\usepackage{longtable}

\usepackage[utf8]{inputenc}

\usepackage{microtype}

\usepackage{inconsolata}
\usepackage{url}
\usepackage{booktabs}
\usepackage{amsfonts}
\usepackage{nicefrac}
\usepackage{microtype}      
\usepackage{xcolor}         
\usepackage[ruled]{algorithm2e}
\SetAlFnt{\small}
\usepackage{amsmath}
\usepackage{bm}
\usepackage{bbold}
\usepackage{graphicx}
\usepackage{multirow}
\usepackage{textcomp}
\usepackage{wrapfig}
\usepackage{xspace}
\usepackage{interval}
\usepackage{bm,upgreek}
\usepackage{colortbl}
\usepackage{enumitem}
\usepackage{pifont}
\usepackage{multirow,makecell}
\usepackage{subfigure}

\newcommand\metricname{SELM\xspace}

\makeatletter
\newcommand*{\@rowstyle}{}
\newcommand*{\rowstyle}[1]{
  \gdef\@rowstyle{#1}%
  \@rowstyle\ignorespaces%
}
\newcolumntype{=}{
  >{\gdef\@rowstyle{}}%
}
\newcolumntype{+}{
  >{\@rowstyle}%
}
\makeatother

\title{SELM: Enhancing Speech Emotion Recognition for Out-of-Domain Scenarios}

\name[affiliation={}]{Hazim}{Bukhari}
\name[affiliation={}]{Soham}{Deshmukh}
\name[affiliation={}]{Hira}{Dhamyal}
\name[affiliation={}]{Bhiksha}{Raj}
\name[affiliation={}]{Rita}{Singh}

\address{
  $ $Carnegie Mellon University
  }
\email{\texttt{hbukhari}@andrew.cmu.edu}


\begin{document}

\setlength{\belowdisplayskip}{4pt} \setlength{\belowdisplayshortskip}{4pt}
\setlength{\abovedisplayskip}{4pt} \setlength{\abovedisplayshortskip}{4pt}

\maketitle


\begin{abstract}Speech Emotion Recognition (SER) has been traditionally formulated as a classification task. However, emotions are generally a spectrum whose distribution varies from situation to situation leading to poor Out-of-Domain (OOD) performance. We take inspiration from statistical formulation of Automatic Speech Recognition (ASR) and formulate the SER task as generating the most likely sequence of text tokens to infer emotion. The formulation breaks SER into predicting acoustic model features weighted by language model prediction. As an instance of this approach, we present \metricname, an audio-conditioned language model for SER that predicts different emotion views. We train \metricname on curated speech emotion corpus and test it on three OOD datasets (RAVDESS, CREMAD, IEMOCAP) not used in training. \metricname achieves significant improvements over the state-of-the-art baselines, with 17\% and 7\% relative accuracy gains for RAVDESS and CREMA-D, respectively. Moreover, \metricname can further boost its performance by Few-Shot Learning using a few annotated examples. The results highlight the effectiveness of our SER formulation, especially to improve performance in OOD scenarios. 
\end{abstract}

\vspace{-0.05in}
\section{Introduction} \vspace{-0.05in}
Understanding emotions from spoken language helps humans in decision-making and social interaction. Speech Emotion Recognition (SER) aims to enable machines to detect emotions from human speech. The emotions are commonly represented as either categorical emotions (\textit{Happy, Angry, Sad}), as Sentiments (\textit{Positive, Neutral, Negative}), or as continuous values of valence, arousal, and dominance ratings. Automatically detecting emotions from speech has multiple applications in various domains such as interactive voice assistants, caller-agent conversation analysis, health care, education, and entertainment. 

Literature on SER formulates it as a classification task \cite{dhamyal2022positional}. The model architecture consists of pretrained speech encoder like Wav2vec2 \cite{wav2vec2}, HuBERT \cite{hubert}, and Whisper \cite{whisper} followed by a classifier. The pretrained speech encoder is usually completely or partially frozen and the classifier is trained from scratch on the target data. This model architecture performs well on in-domain data \cite{saliba2024layer}. However, the performance drops drastically on out-of-domain data. Moreover, the learned classifier is static and only predicts predefined emotions and hence needs to be retrained for every domain and target task. The recent works in audio-text learning \cite{pengi, elizalde2022clap,gong2023listen,gong2023joint, deshmukh23_interspeech} aim to avoid training classifiers for each target task and domain. To remove the fixed classifier stage, audio-text models use cosine similarity between audio embeddings and classes to determine predictions where the classes are represented by text embeddings. The popular instances of such approaches like CLAP \cite{elizalde2022clap, elizalde2024natural, wu2023large}, Pengi \cite{pengi}, LTU \cite{gong2023listen,gong2023joint} allow users to define emotion classes to predict at test-time. Moreover, as emotion is a spectrum, flexible class prediction enables users to define emotion as “happy but little sad” and “initially happy tone but overall sad” or “happy initially and sad later”. Though audio-text approaches are trained on large-scale audio data, they still cannot generalize to OOD scenarios, where the speech data differs from the training data in terms of speakers, languages, accents, recording conditions, or emotional categories. For example, \textit{Happy} emotion in call-center has a significantly different energy than \textit{Happy} in business meetings, leading to misclassifications. This reduces the applicability of audio-text models for SER in-the-wild. 

Few-Shot Learning (FSL) approaches utilize a few annotated examples from the target domain to adapt the pre-trained model. This allows SER models to be adapted to different domains like call-center, and health care with few examples and minimal updates. Two popular FSL approaches for SER are LanSER \cite{gong2023lanser} and AudioFlamingo \cite{audioflamingo}. LaNSER utilises a pre-trained Large Language Model (LLM) through weakly-supervised learning and then uses 10\% of data to improve OOD performance. Similarly, AudioFlamingo takes the approach of building a strong foundation audio model that can be later adapted to OOD dataset using as less as 4-shots. However, there exists a gap in in-domain performance and the OOD performance of AudioFlamingo and LanSER. Moreover, LaNSER requires large amount of annotated audio data to improve OOD performance. For example, 10\% for CREMAD \cite{cao2014crema} is 744 audio files, which is costly to annotate. 

In this work, we provide an alternate formulation for SER and propose \metricname, a speech emotion recognition model that improves OOD performance and is adaptable to new test distribution with minimal or few-annotated examples. SELM uses an audio-conditioned language modeling approach for SER where the prediction of different emotion views is formulated as a text generation task. Our contributions are as follows: (1) An alternate formulation of the task of SER which breaks SER into Acoustic Model and Language Model. (2) Introduce \metricname which is an instance of the above formulation and provides SoTA performance in Out-of-Domain scenarios. (3) Propose Few-Shot Learning approach for \metricname (4) Extensively testing \metricname in various setups: In-domain, OOD, and Few-Shot Learning on three public datasets (RAVDESS, CREMA-D and IEMOCAP), thus establishing a baseline for future work. 
\section{Formulation} \label{sec: formulation}
An ASR system produces the most likely word sequence for a speech signal. The statistical approach to ASR has been:
$$w^* = arg\max_{w} p(x|w) p(w)$$
where $x$ represents the audio, $w$ represents a word sequence and the optimal word decoding is $w^*$. The $p(w)$ is modeled by the language model and $p(x|w)$ is calculated by the acoustic model. 

\noindent Traditionally speech emotion recognition has been formulated as a classification problem. However, one can formulate speech emotion recognition as generating emotion tokens ($e$), conditioned on audio, where the emotion tokens are actually text tokens describing emotion. Taking inspiration from the ASR formulation, we provide the statistical formulation for speech emotion recognition as :
$$e^* = arg\max_{e} p(e|x,w)$$
where we want to predict the best emotion sequence given acoustic sequence $x$ and word sequence $w$ (transcript). The emotion is considered a sequence of tokens instead of classes. In practice, we use Byte pair encoding used by GPT2 to convert text into tokens. Following Bayes’ rule, the posterior probability in the above equation can be expressed as:
$$e^* = arg\max_{e} \frac{p(x|e,w)p(e,w)}{p(x,w)}$$
$$e^* = arg\max_{e} \frac{p(x|e,w)p(e|w)p(w)}{p(x|w)p(w)}$$
$$e^* = arg\max_{e} \frac{p(x|e,w)p(e|w)}{p(x|w)}$$
$p(x|w)$ is constant with respect to ranking of hypothesis for $e$, hence can be discarded
$$e^* = arg\max_{e} p(x|e,w)p(e|w)$$ 
The $p(e|w)$ tells us given the word sequence what is the likely emotion. This can be modeled by a pre-trained Language Model, for example GPT2. The $p(x|e,w)$ tells us what acoustic features can be observed given the particular emotion and word sequence (transcript). It should contain a representation of the distinct sounds that make up each emotion and word in language model.
\section{Methodology}
\begin{figure}[!ht]
   \centering    \includegraphics[width=0.43\textwidth]{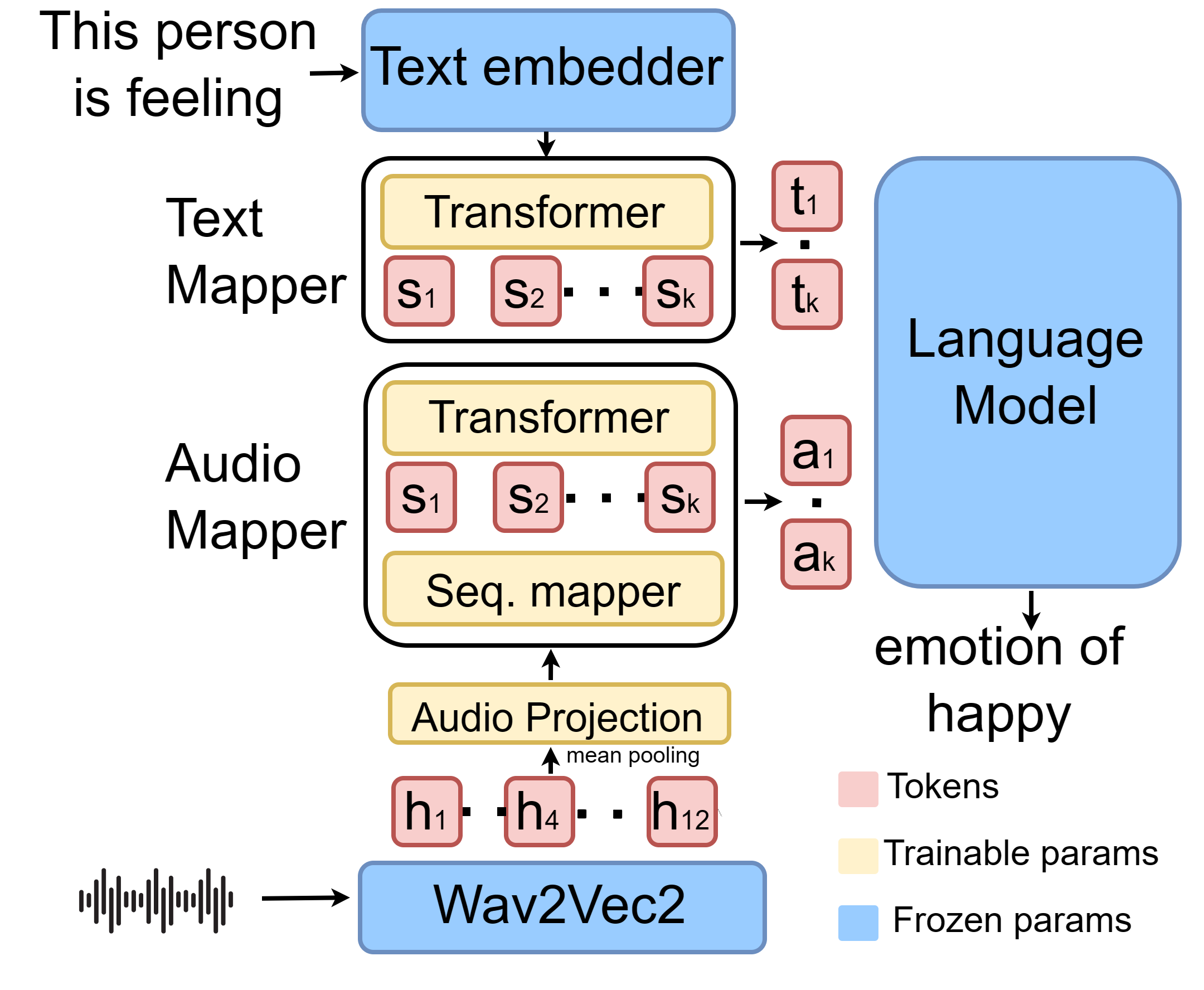}
     \caption{\metricname: \underline{S}peech \underline{E}motion \underline{L}anguage \underline{M}odel. The model is fed with audio and input prompts, which get independently encoded by audio projection \& audio mapper, and text embedder. The encoded audio and text is used to prompt a Language Model. In the figure, the input text prompt is \textit{``this person is feeling"} and \metricname outputs \textit{``emotion of happy"}. The audio projection and audio mapper are learned during training while the Language Model and Wav2vec2 are frozen.\vspace{-0.3 in}}
     \label{fig:SELM}
\end{figure}

We realize the formulation (Section \ref{sec: formulation}) as audio-conditioned language modeling for Speech Emotion Recognition. In this section, we describe \metricname (Figure \ref{fig:SELM}), which takes an audio recording and prompt as input and produces text as output.

\subsection{Architecture} \label{sec: Architecture}

\noindent \textbf{Audio encoder.} The audio encoder extracts dense representations from audio. We use Wav2Vec2 as the choice of audio encoder due to its near SOTA performance for SER \cite{saliba2024layer}. The 4th layer of Wav2Vec2 contains the information relevant to the task of SER \cite{saliba2024layer}. Therefore, we extract the 4th layer of Wav2vec2 as our frozen audio feature extractor. 

\noindent \textbf{Text Embedder.} The text embedder consists of a Language Model tokenizer followed by embedding lookup to convert language model tokens to continuous embeddings. We keep the tokenizer and embedding lookup same as the Language Model used, which in our case is GPT2 \cite{radford2019language}

\noindent \textbf{Audio Projection.} The audio projection consists of two learnable linear layers with GeLU activation in between. This is similar to CLAP projection \cite{elizalde2022clap}, and performs a non-linear transformation of the 4th layer of Wav2Vec2 hidden state. This transformed representation is passed to the Audio Mapper. 

\noindent \textbf{Audio Mapper.} The audio mapper consists of a sequence mapper and a transformer layer. The sequence mapper is a linear layer followed by a reshape operation to convert a single embedding into a fixed set of continuous embeddings ($t_1$, $t_2$, ... $t_k$). The sequence is passed to a self-attention Transformer. The transformer learns to map the generated sequence from audio ($s_1$,... $s_k$) to the latent space of the Language Model ($a_1$,... $a_k$). 

\noindent \textbf{Text Mapper.} The text mapper consists of a learnable Transformer layer to transform text embeddings into the latent space of the Language Model ($t_1$, $t_2$, ... $t_k$). 

\noindent \textbf{Language Model.} The latent prefix for the Language Model is the concatenation of the audio prefix ($a_1$, $a_2$, ... $a_k$) and text prefix ($t_1$, $t_2$, ... $t_k$). The Language Model generates a text (\textit{emotion of happy}) conditioned on this latent prefix. We use GPT2 \cite{radford2019language} as the choice of Language Model and a prefix size of 10 for audio and text respectively. 

\subsection{Training} \label{subsec: training}
The model is trained using the next-token prediction objective. Let one example of training data consist of audio $x_{a_i}$, input text prompt $x_{t_i}$ and ground truth response $y_{t_i}$. The audio encoder and audio mapper transform the audio $x_{a_i}$ into a continuos sequence of embeddings $(a_{1_i}, a_{2_i}, .... a_{k_i})$, where k is the prefix size. Simlarly, the text embedder and text mapper transforms the prompt $x_{t_i}$ into a continuos sequence of embeddings $t_{1_i}, t_{2_i}, .... t_{k_i}$. Both sequences are concatenated to form the prefix $z_i$: 
\begin{equation}
    z_i = a_{1_i}, a_{2_i}, .... a_{k_i},t_{1_i}, t_{2_i}, .... t_{k_i} \label{equation: prefix} 
\end{equation}
Then the Language Model conditioned on $z_i$ produces a sequence of output tokens $o_{t_i}$. The model is trained to predict the next text tokens $o_{t_i}$ ($t \in [0,l]$) conditioned on $z_i$ in an autoregressive fashion. The loss function is Cross-Entropy:
\begin{equation}
\mathcal{L} = - \sum_{t=1}^{l} \log p_{\theta} (y_{t_i} | z_i, y_{0_i},...,y_{l_i}) 
\end{equation}
where $\theta$ denotes the model's trainable parameters which consist of Audio Projection, Audio Mapper, and Text Mapper. We keep the Wav2Vec2, Text embedder, and Language Model frozen.

\subsection{Inference} \label{subsec: inference}
The test audio and test input prompt are used to build prefix $z_i$ to prompt the Language Model. The Language Model predicts the next token based on beam search decoding with a beam size of 3. The output of the model is a text and can be directly used in any downstream pipeline

For evaluating metrics, the free-form text answer from SELM has to be converted to 1 out of C classes for each dataset or user-specified classes. For example, \metricname generates the emotion as frustration but the user wants classification into 4 classes of \textit{Happy, Sad, Angry, Neutral}. Therefore, the generated text needs to be mapped into one of the C classes. For this, we encode the generated text and classes with CLAPS text encoder. Then use cosine similarity between text embedding of generated text and text embedding of classes to determine the most likely class. 

\section{Experimental setup} \label{sec:exp}

\subsection{Datasets}
\noindent \textbf{Training Datasets:} 
The training dataset for \metricname consists of triplets of form (audio, prompt, output). Training on all emotion views is proven to improve performance \cite{10096700}. Hence, we train \metricname to predict all three views of emotions: categorical, sentiment, and dimensional. For categorical data, we source audio and labels from MSP Podcast \cite{msp_podcast}, CMU MOSI \cite{zadeh2016mosi}, CMU MOSEI \cite{mosei}, MELD \cite{poria2019meld} and use "This person is" as the prompt. For sentiment data, we convert categorical labels from MSP Podcast and MELD to sentiment with "This sentiment is" as a prompt. While, for dimensional scores, we utilize the labels available from the MSP Podcast and prompt with "Describe emotion parameters". In total, we train on 315k triplets.

\begin{table}[!ht]
\footnotesize
\center
\begin{tabular}{cccccccc}\hline
Dataset & Files & Dur. (secs) & Output & Setup \\ \hline
RAVDESS & ~2.5k & $\leq$ 5 & MC (8) & 5 folds \\ 
CREMA-D & ~7k & 5 & MC (6) & 5 folds \\
IEMOCAP & 2 & 4.5 & MC (6) & train/test \\ \hline
\end{tabular}
\caption{\label{table: downstream tasks}
The datasets used to simulate OOD setup. They differ in terms of speakers, accents, recording conditions, and emotional categories. For the ``Output Type" column, MC to multiclass and (.) represents the classes. \vspace{-0.2in}
}
\end{table}

\noindent \textbf{Evaluation Datasets:} 
We use three datasets for evaluating SER performance: RAVDESS \cite{ravdess}, CREMAD \cite{cao2014crema}, IEMOCAP \cite{busso2008iemocap}. These datasets differ in the number of emotional classes, the type of features (audio-only vs. multimodal), and their specific data collection setup (Table \ref{table: downstream tasks}). The above datasets are not used in training SELM. This makes the three datasets ideal for simulating OOD setup and for benchmarking.

\subsection{Setup} \label{sec: setup}
\noindent \textbf{In-Domain.} To check the in-distribution performance of \metricname we use the in-domain setup. In this setup, the model is trained using the training subset of the Evaluation Datasets and evaluated on the testing subset of them. In explicit train-test subset are not available, we perform N-fold cross-validation.

\noindent \textbf{Out-of-Domain.} We simulate the Out-of-Distribution setup by excluding some datasets from training. Therefore, the model is evaluated on three Evaluation Datasets namely RAVDESS, CREMA-D, and IEMOCAP, not used in training. 

\noindent \textbf{Few-Shot Learning.} Adapting to new distribution requires annotated examples for the target domain. To test the efficiency of this adaptation, we explore Few-Shot Learning (FSL) for \metricname.  
For the FSL setup, we inherit the model pretrained \metricname model. Then the model is finetuned using N examples from target domain. The finetuning is restricted to specific parts of model to be parameter efficient and to prevent loss of generalization ability of \metricname. We perform FSL in two settings, 4-shot and 8-shot, where 4 examples are 8 examples per class are randomly picked from the training set to finetune specific parameters of \metricname.

\section{Results}
Unweighted accuracy is used as the primary evaluation metric. The experiments in this section are designed to test \metricname in different scenarios, namely: In-domain (Section \ref{sec: In-domain}), Out-of-Domain (Section \ref{sec:OOD performance}), Few-Shot (Section \ref{sec: Few-shot learning}). The Section \ref{sec:Ablation on Few-Shot Learning} contains ablation studies to determine which parameters to update for Few-Shot Learning.

\subsection{In-Domain} \label{sec: In-domain}
In this Section, we compare our formulation of SER with the traditional classification-based SER. In \metricname, the audio encoder is Wav2Vec and is frozen i.e. not learned during training. We compare against literature baselines where the Wav2Vec2 is frozen followed by learning single or multiple linear layers. This experiment is called In-Domain setup as the model is trained on the train subset of the dataset followed by testing on the test subset. The results are shown in Table \ref{tab:in-domain-result} for three datasets. We present \metricname performance in the first row, and compare it to the methodology where Wav2Vec2 features are used followed by a simple neural network on top. We believe this is a fair comparison as the acoustic model architecture is the same in both models. \metricname's better performance shows the benefit of decomposing the SER task into an Acoustic Model followed by a Language Model reweighting.

\begin{table}[tbh!]
    \footnotesize
    \centering
    \begin{tabular}{c|ccc} \hline
         & \multicolumn{3}{c}{In-domain Dataset} \\ \hline
        Model & RAVDESS  $\uparrow$ & CREMAD $\uparrow$ & IEMOCAP  $\uparrow$ \\ \hline  
        Wav2Vec2 FT & 56.53 \cite{luna2021proposal} & 46.02 \cite{phukan2023comparative} & 69.90 \cite{chen2023exploring}\\
        SELM & 75.70 & 88.16 & 73.09 \\ \hline
    \end{tabular}
    \caption{In-domain performance of \metricname on three datasets. Similar to \metricname, the benchmark numbers also use wav2vec2-base embeddings to extract acoustic features. 
    \vspace{-0.3in}}
    \label{tab:in-domain-result}
\end{table}

\subsection{Out-of-Domain} \label{sec:OOD performance}
In this Section, we compare \metricname's Out-of-Domain (OOD) performance against benchmark models from Literature. The existing literature models are also not trained on these three datasets, making this a valid OOD comparison. Table \ref{tab:ood-domain-result} presents the OOD experimental results for RAVDESS, CREAM-D and IEMOCAP. The first half of the table presents performance on all emotion classes in the dataset, while the lower half (denoted with *) presents numbers on a subset of classes. The subset of classes are chosen to be the four primary emotions: \textit{happy, sad, angry, neutral}. The results show \metricname outperforms existing models on all three datasets for both all-class and 4-class setup. Moreover, some of the datasets contain emotions that \metricname has not seen during training, and hence showcasing the generalization ability of the model. 

\begin{table}[tbh!]
\footnotesize
\center
\begin{tabular}{c|ccc} \hline
& \multicolumn{3}{c}{Out-of-Domain Dataset} \\ \hline
Model & RAVD $\uparrow$ & CREMA $\uparrow$ & IEMOC $\uparrow$ \\ \hline
Random & 12.50 & 16.70 & - \\
CLAP \cite{elizalde2022clap} & 16.00 & 17.80 & 13.71 \\
Pengi \cite{pengi} & 20.32 & 18.46 & - \\
MMS large \cite{ma2023investigating} & 13.50 & 17.20 & - \\
Whisper medium.en \cite{ma2023investigating} & 15.30 & 20.90 & - \\
Whisper medium \cite{ma2023investigating} & 16.70 & 19.90 & - \\ 
Whisper large-v2 \cite{ma2023investigating} & 15.10 & 20.20 & - \\ 
AudioFlamingo \cite{audioflamingo} & 20.90 & 26.50 & - \\
\metricname & \textbf{24.51} & \textbf{28.30} & \textbf{21.42} \\ \hline
CLAP $^*$ \cite{elizalde2022clap} & 29.48 & 31.05 & 33.72 \\
LanSER PS$^*$ \cite{gong2023lanser} & - & 15.90 & 30.90 \\
LanSER CM$^*$ \cite{gong2023lanser} & - & 23.50 & 34.30 \\ 
EmoCLAP$^*$ \cite{dhamyal2022describing} & 38.46 & 35.22 & - \\
\metricname$^*$ & \textbf{52.53} & \textbf{42.79} & \textbf{40.02} \\ \hline
\end{tabular}
\caption{\label{tab:ood-domain-result}
OOD Performance of different models across three datasets. The dataset is considered OOD when labeled or unlabelled audio is not used during training or unsupervised adaptation during testing. The $^*$ symbol indicates only four emotion classes (anger, happiness, sadness, and neutral) are used for evaluation. The metric used is unweighted.  \vspace{-0.4in}
}
\end{table}

\subsection{Few-Shot Learning} \label{sec: Few-shot learning}
In this Section, we compare adaptation \cite{deshmukh2024domain} like Few-Shot learning performance of \metricname against literature benchmarks. The results are shown in Table \ref{tab:few-shot-result}. The table shows the performance of the model when it is finetuned on either 4 examples per class (4-shot) or 8 examples per class (8-shot). For the all-class settings (presented in the earlier rows of the table), we compare \metricname's performance against Audio-Flamingo \cite{audioflamingo}. The AudioFlamingo reports 4-shot and 8-shot numbers, however the testing strategy and split is not reported. For the 4 class emotion classification, we compare againts LanSER \cite{gong2023lanser}, which reports the performance when the model is fine-tuned using 10\% of the data. We note that 10\% data from target domain is significantly more than 4-shot. For example, 10\% of data from CREMA-D is equivalent to 744 audio files and hence the setups are not directly comparable. Due to the lack of Few-Shot Learning approaches in SER, we compare \metricname against LanSER

The results shown in Table \ref{tab:few-shot-result} lead to multiple conclusions. First, \metricname performs better than prior work on CREMA-D and IEMOCAP. Second, 8-shot always leads to performance improvement over 4 shot. Similarly, 16-shot performs better than 8-shots, i.e. providing \metricname more target domain data will lead to better performance. However, there \metricname Few-Shot has some limitations. First, the AudioFlamingo performs better on RAVDESS in all-class settings than \metricname. We believe that this is due to the specific training strategy of AudioFlamingo which enables use of In-Context Learning. Moreover, AudioFlamingo is trained on 6 million instances, which is 20 times higher than our settings. Second, in RAVDESS, the 4-shot leads to higher performance than 8-shot in the 4-class settings. This might be because RAVDESS has song and speech samples, while \metricname has never encountered songs in training data. As songs have different audio distribution from speech samples, the parameter-efficient Few-Shot learning is not sufficient to improve performance and adapt to the song distribution in a few examples.

\begin{table}[tbh!]
\footnotesize
\center
\begin{tabular}{c|c|ccc}\hline
&  & \multicolumn{3}{c}{Dataset for Few-Shot Learning} \\ \hline
Model & Setup & RAVD  $\uparrow$ & CREMA $\uparrow$ & IEMOC  $\uparrow$ \\ \hline
AudioFlam. & 4-shot & - & 30.47  & - \\
AudioFlam. & 8-shot & \textbf{35.20} & 31.80  & - \\
\metricname & 4-shot & 30.09 & 30.10 & 25.32 \\
\metricname & 8-shot & 31.81 & \textbf{32.27} & \textbf{27.01} \\ \hline
LanSER PS$^*$ & 10\% & - & 35.50 & 42.00 \\
LanSER CM$^*$ & 10\% & - & 43.70 & 50.00 \\
\metricname$^*$ & 4-shot & \textbf{57.14} & 43.93 & 50.17 \\
\metricname$^*$ & 8-shot & 55.77 & \textbf{46.02} & \textbf{50.17} \\ \hline
\end{tabular}
\caption{\label{tab:few-shot-result}
Few-Shot Learning results of different models and methods on OOD dataset. The dataset is considered OOD when labeled or unlabelled audio is not used during training or unsupervised adaptation during testing. The $^*$ symbol indicates only four emotion classes (anger, happiness, sadness, and neutral) are used for evaluation. The metric used is unweighted accuracy.  
} \vspace{-0.4in}
\end{table}

\vspace{-0.1in}
\subsection{Ablation on Few-Shot Learning} \label{sec:Ablation on Few-Shot Learning} 
\begin{figure}[!htb]
    \centering    \subfigure{\includegraphics[height=0.15\textheight,width=0.38\textwidth]{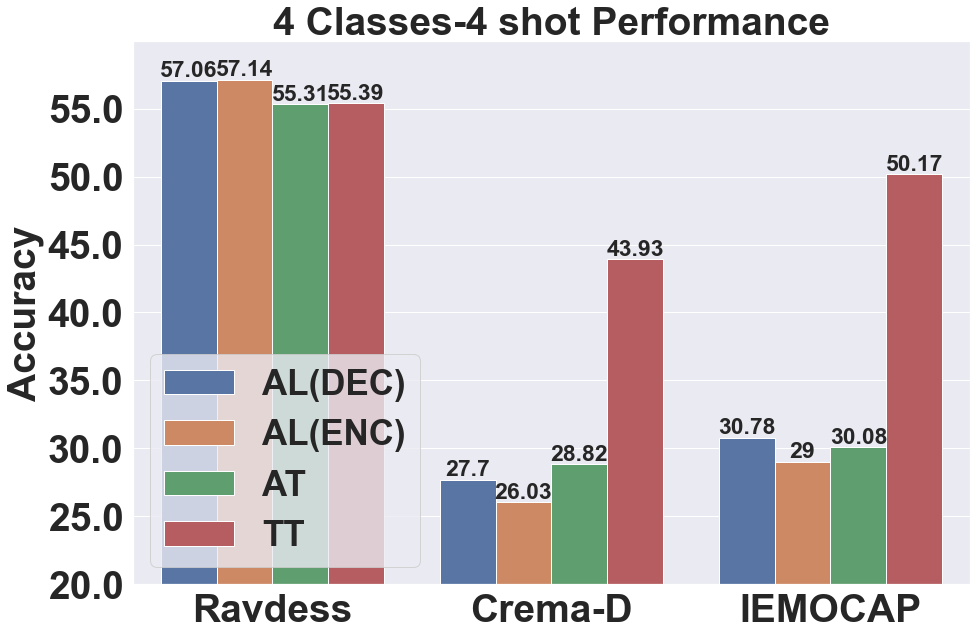}}  \subfigure{\includegraphics[height=0.15\textheight,width=0.38\textwidth]{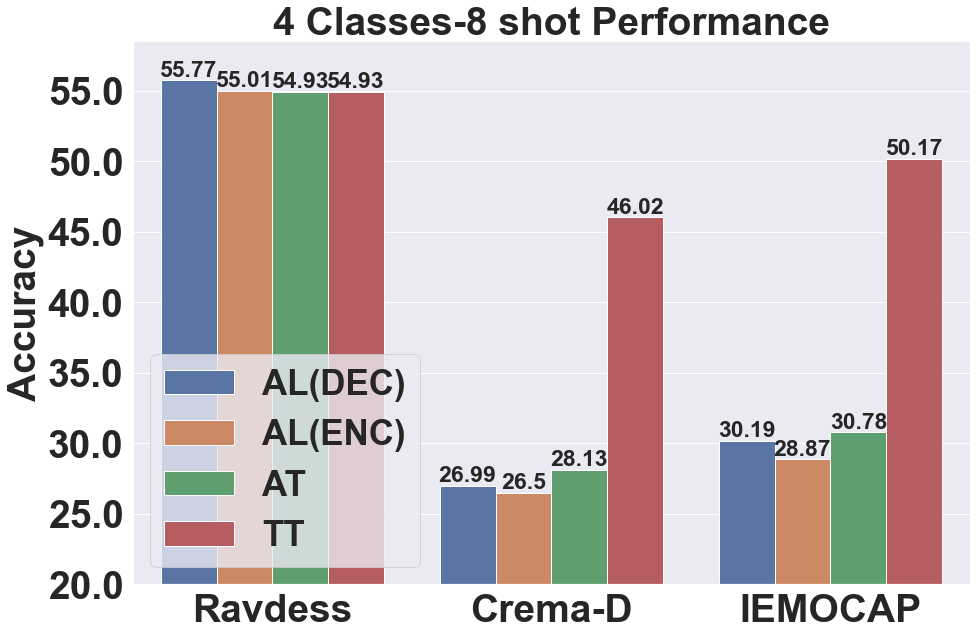}}
    \vspace{-0.15in}
    \caption{Ablation study on the parameters to update for Few-Shot Learning. The first graph and second graph show performance improvement achieved by finetuning different parts of the model under 4-shot and 8-shot settings for the three datasets. \vspace{-0.25in}}
    \label{fig:PCC T2A}
\end{figure}

\vspace{-0.05in}
To understand which parameter of \metricname to update for Few-Shot Learning, we perform multiple ablation studies. The parameters under consideration are: linear layer that is part of the decoder (AL-Dec) or the linear layer part of the encoder (AL-Enc), or the Audio Mapper Transformer (AT) or the Text Mapper Transformer (TT). The results of the ablation study are shown in Figure \ref{fig:PCC T2A}. We find that finetuning Text mapper from \metricname leads to the best performance improvement when data distribution is similar. However, for completely different data distribution, adapting a parameter in audio mapper performs better. For example, RAVDESS contains songs which \metricname has not seen in training data. To adapt to a widely different audio distribution, requires training parameters in the audio mapper.

\section{Limitations} \label{sec: limitations} \vspace{-0.05in}
\metricname achieves SoTA Out-of-Domain performance. Moreover, by using Few-Shot Learning, the OOD performance can be further improved. However, \metricname has limitations. First, \metricname is trained only on spoken English datasets and hence its speech emotion recognition will drop on other languages. Second, the Few-Shot Learning improves \metricname's performance on OOD data. However, the improvements are minor when the audio distribution is significantly different.

\vspace{-0.05in}
\section{Conclusion} \label{sec: conclusion} \vspace{-0.05in}
In this work, we provide an alternate formulation for Speech Emotion Recognition, posing it as a text token generation task, consisting of both acoustic and language models. We build \metricname based on this formulation and test on multiple SER datasets. The results show that \metricname generalizes better than the literature models on Out-of-Domain and Few-Shot scenarios. In Out-of-Domain settings, \metricname achieves 17\% and 7\% relative accuracy improvements in RAVDESS and CREAMA-D respectively. In Few-Shot setting (4-shot and 8-shot), on average \metricname achieves better performance over baseline models.

\newpage
\bibliographystyle{IEEEtran}
\bibliography{mybib}

\end{document}